# Reducing Temperature Swing and Rectifying Radiative Heat Transfer for Passive Dynamic Space Thermal Control with Variable-Emittance Coatings

Liping Wang,* Neal Boman, Sydney Taylor, and Chloe Stoops

School for Engineering of Matter, Energy and Transport, Arizona State University, Tempe, Arizona 85287

*Corresponding author: liping.wang@asu.edu*

**Abstract**

Dynamic radiative thermal control is crucial for normal operation and energy saving of spacecraft that copes with changing thermal environment involving heat dissipation to cold deep space, external heating from the Sun and nearby planet, and internal heating from onboard electronics. Variable-emittance coatings, whose infrared emittance can be tuned passively by temperature or actively by external stimuli, could provide a viable solution. In this work, we experimentally demonstrate self-adaptive dynamic radiative heat transfer with variable-emittance coating based on thermochromic $VO_2$ in space-like thermal environment with a coldfinger and a custom-made sample mount inside a vacuum cryostat. Black Actar and highly reflective tungsten mirror are used to calibrate the parasitic head load and heat flux sensor sensitivity, while multiple static-emittance samples made of silicon wafers with different doping levels are measured for validation of the experimental method and for direct comparison with the variable-emittance $VO_2$ coating. With the coldfinger at 80 K to mimic external radiative scenarios in space, the tunable coating exhibits 6-fold enhancement in radiative thermal conductance upon $VO_2$ phase transition for promoted heat dissipation, in addition to reduced temperature swing by almost 20°C compared to the static emitters. With the coldfinger at 25°C as internal radiative scenarios in space, similar 6-fold heat dissipation from the variable-emittance coating is also observed, while radiative heat transfer is much suppressed with a constant radiative thermal conductance when the coldfinger is hotter than the tunable coating at 25°C, leading to a thermal rectification factor of 1.8±0.2 experimentally achieved.

Space objects such as spacecraft, satellites, and probes could receive a wide range of heat loads both externally and internally due to changing environment in space, leading to large temperature swing which could disrupt normal operation and drain limited on-board energy for active thermal control.[1,2] Passive dynamic radiative heat transfer with tunable variable-emittance coatings could provide a viable solution or mitigation to such a challenge.[3-5] As depicted in Fig. 1(a), when the space object is heated externally by different amount when facing or being away from the Sun, a variable-emittance coating whose emittance increases with higher temperature could maintain a constant surface temperature with thermal homeostasis effect.[6] Internally as shown in Fig. 1(b), when the payload is hotter than the surface, augmented heat dissipation is wanted. On the other hand, radiative heat transfer should be suppressed if surface temperature is too high for thermally protecting internal payload and saving energy for active cooling. This could be achieved by a variable-emittance coating with thermal rectification effect.[7]

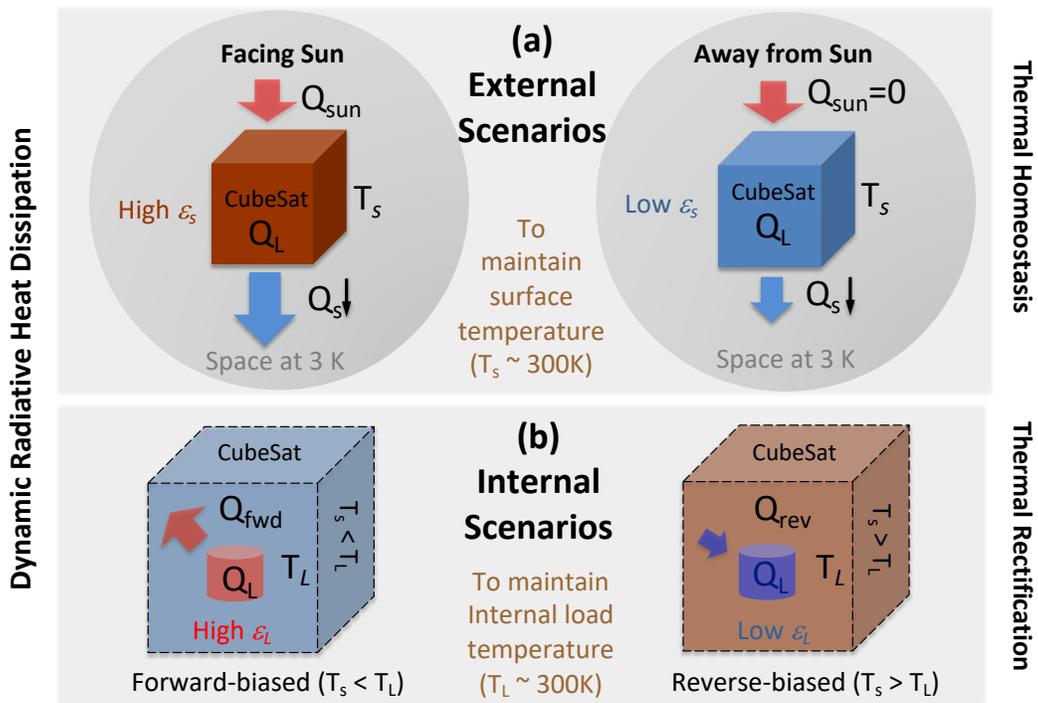

**Fig. 1.** Radiative heat transfer scenarios in space with changing environment: (a) external heat exchange of the spacecraft facing or away from the Sun to maintain surface temperature ($T_s$ ~ 300 K) where thermal homeostasis is desired; and (b) internal heat exchange between the spacecraft and heat load to maintain the load temperature ($T_L$ ~ 300 K) where thermal rectification is desired. In both scenarios, dynamic radiative heat dissipation is required.

Passive variable-emittance coatings are usually made of thermochromic materials such as lanthanum strontium manganese oxide (LSMO)[8,9] and vanadium dioxide ($VO_2$)[10,11] by taking advantage of their unique phase transition behaviors based on temperature change without any external input. $VO_2$ has attracted most of recent research attentions as its insulator-to-metal phase transition (IMT), which is much broader over 275°C for LSMO,[12] could occur within a narrow 20°C temperature range around 68°C. However, $VO_2$ cannot be used directly as it is reflective in its metallic phase at high temperatures, where high emissivity is desired for enhanced heat dissipation. Recently, nanophotonics has been adopted to develop $VO_2$-based variable-emittance coatings. Various of planar metafilm structures[13-21] in typical $VO_2$-dielectric-metal thin film stacks, which could achieve wavelength-selective Fabry-Perot (FP) cavity resonant absorption with only metallic $VO_2$ phase, have been theoretically designed[13] and optimized[14,15]. Large emittance change from these planar tunable coatings has been experimentally demonstrated.[16-21] Micro/nanostructured photonic $VO_2$ coatings[22-27] with variable emittance based on different physical mechanisms are also widely studied both theoretically and experimentally, while it is more challenging to scale up due to costly fabrication processes compared to their planar counterparts. The performance in the adaptive thermal control with variable-emittance coatings, the ability in reducing temperature swing, and energy-saving benefit have been comprehensively evaluated theoretically,[28-30] while very few studies[31,32] have been reported to experimentally demonstrate the dynamic thermal control with these variable-emittance coatings in space-like thermal environment.

In this work, we experimentally demonstrate reduced temperature swing with thermal homeostasis effect and rectified radiative heat transfer with $VO_2$ based thermochromic variable-emittance coatings in space-like cryothermal environment, which is realized with a temperature-controlled coldfinger from 80 K to 25°C and a custom-made sample mount inside a vacuum cryostat. Tunable $VO_2$ based Fabry-Perot (VO2FP) variable-emittance coating made of 55-nm $VO_2$, 500-nm silicon and 200-nm aluminum thin films was fabricated on a double-side polished silicon wafer via thin film sputtering and furnace oxidation following our previous work.[20] It is expected to achieve high emittance at temperatures above 68°C where Fabry-Perot resonance is excited with metallic $VO_2$, while it remains highly reflective with insulating $VO_2$ at low temperatures. A Fourier transform infrared spectrometer (Thermo Fisher Scientific, iS50) along with a variable-angle reflection accessory (Harrick Scientific, Seagull) was used to measure the spectral specular reflectance at 10° incidence angle in the wavelength range from 2 μm to 22 μm at a resolution of 4 cm$^{-1}$ with each spectrum averaged over 32 scans. A freshly deposited aluminum mirror was used as the reference, while the spectral reflectance of samples was corrected with the theoretical reflectance of aluminum whose optical constants were obtained from Palik.[33] A home-made temperature stage was used to control the sample temperature with every 4°C increment from room temperature to 100°C. Spectral measurement was taken after each temperature reached the setpoint for 5 mins with fluctuation less than 1°C.

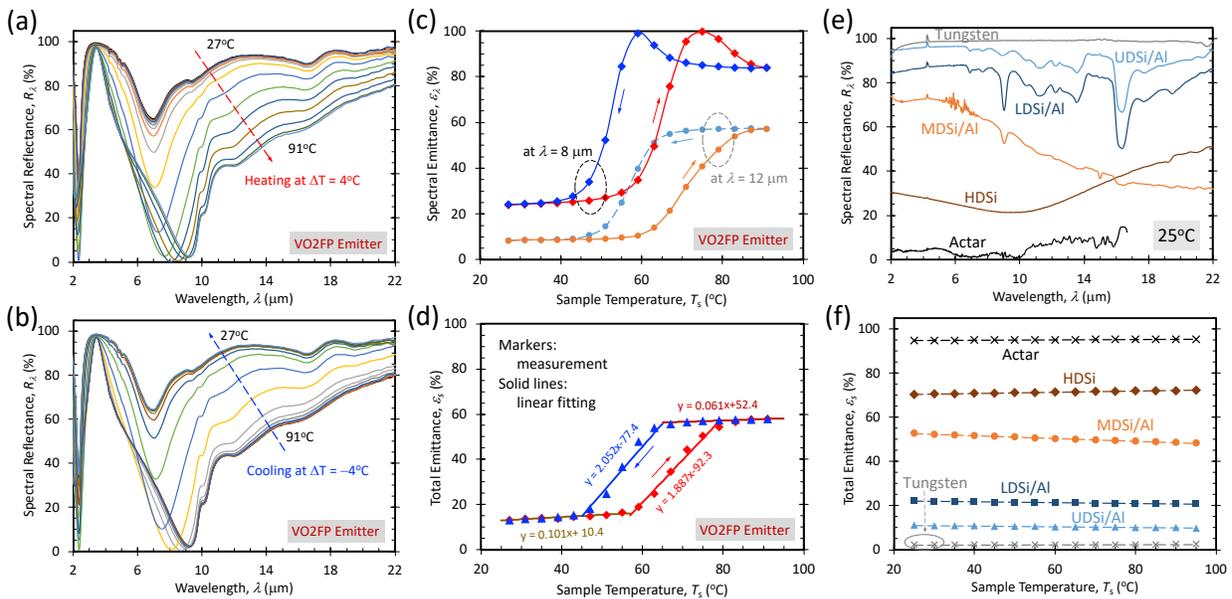

**Fig. 2.** Measured temperature-dependent spectral infrared reflectance of tunable VO2FP emitter between 27°C to 91°C upon (a) heating and (b) cooling; (c) measured spectral emittance at selected wavelengths ($\lambda$ = 8 and 12 μm) and (d) total normal emittance of the tunable VO2FP emitter as a function of temperature; (e) measured spectral infrared reflectance at 25°C and (f) total emittance at different temperatures of reference samples including tungsten mirror, undoped silicon with Al backside (UDSi/Al), lightly doped silicon with Al backside (LDSi/Al), medium doped silicon with Al backside (MDSi/Al), heavily doped silicon (HDSi), and black Actar.

Figs. 2(a) and 2(b) show the measured temperature-dependent spectral infrared reflectance $R_\lambda$ of the tunable VO2FP emitter between 27°C to 91°C upon heating and cooling, respectively. Variation of spectral reflectance with temperature is clearly observed. In particular, a low reflectance dip around $\lambda$ = 8 µm wavelength with increasing temperature appears due to the excitation of FP resonance once VO$_2$ becomes metallic. The temperature-dependent spectral emittance $\varepsilon_\lambda$ of the opaque VO2FP sample can be simply obtained by $1 - R_\lambda$, as shown in Fig. 2(c) at two selected wavelengths. At $\lambda$ = 12 µm wavelength, the spectral emittance exhibits typical IMT and hysteresis behaviors of VO$_2$ with monotonic increase with higher temperatures during the phase transition, leading to the change from 0.08 at 27°C to 0.58 at 91°C. However, at $\lambda$ = 8 µm wavelength where the FP resonance is excited, the spectra emittance starts with 0.25 at room temperature, achieves a peak of unity at 75°C upon heating (or at 59°C upon cooling), and stabilizes at 0.84 beyond 91°C.

For the evaluation of radiative heat transfer, total emittance of the tunable VO2FP emitter is calculated as $\varepsilon_s(T_s) = \int_0^\infty \varepsilon_\lambda(T_s) E_{B,\lambda}(T_s) d\lambda / \sigma T_s^4$, where $E_{B,\lambda}$ is the spectral blackbody emissive power from Planck's law.[34] As shown in Fig. 2(d), the total normal emittance of the tunable VO2FP emitter remains almost the same around 0.12 at low temperatures with insulating VO$_2$ phase. Upon phase transition, the total emittance increases quickly and finally reaches around 0.57 with metallic VO$_2$. Besides, a thermal hysteresis about 12°C can be observed between heating and cooling processes. To facilitate the theoretical modeling, the measured total normal emittance of the tunable VO2FP emitter is fitted with four linear relations for its insulating phase, transition during heating, transition during cooling and metallic phase, as labelled in the figure. Our previous theoretical calculation[13] suggested slight angular dependance of the total emittance for the VO2FP emitter at metallic phase due to the nature of wave interference when FP resonance is excited. Therefore, the total hemispherical emittance of the VO2FP emitter in the metallic phase is reduced by 10% from total normal emittance, while it remains the same in the insulating phase without excitation of FP resonance.

For calibration and validation of the cryothermal tests, several static-emittance samples were prepared with a wide range of emittance values, including tungsten mirror, black Actar, as well as undoped, lightly-doped, medium-doped, and heavily-doped silicon wafers. Tungsten mirror was fabricated by sputtering 200-nm tungsten at 0.15 nm/s onto a polished silicon wafer. Double-side polished silicon wafers of 280 µm thick with various doping levels were commercially purchased with different resistivities. To ensure the infrared opaqueness, 200-nm aluminum was sputtered on the backside of undoped, lightly and medium doped silicon samples. Commercial black Actar sample was attached to a polished silicon wafer with thermal paste.

The spectral reflectance of these static-emittance samples was measured at 25°C as shown in Fig. 2(e). Note that black Actar was measured with a gold integrating sphere (PIKE Technologies, Mid-IR IntegratIR) due to its diffuse surface, while all other samples with smooth surfaces were measured for specular reflectance at near-normal incidence. Also, it is known that the infrared properties of doped silicon could vary with temperature. However, within the small temperature range (i.e., 20°C to 100°C) tested here, the change in the spectral reflectance of doped silicon is less than 3% (see Fig. S1 in the Supplementary Material for the measured spectral reflectance of doped silicon samples at different temperatures), which is considered negligible. Fig. 2(f) presents the total emittance from 20°C to 100°C calculated from the measured spectral reflectance for these calibration and validation samples. Black Actar has the highest total emittance of 0.951±0.004, and the tungsten mirror has the lowest constant value of 0.025±0.002. The total emittance is 0.105±0.007, 0.213±0.008, 0.505±0.025, and 0.714±0.011 for undoped, lightly, medium and heavily doped silicon wafers as validation samples. Note that, the undoped and medium doped silicon samples have almost the same total hemispherical emittance with the tunable VO2FP emitter in its insulating and metallic phase, respectively.

The radiative thermal tests were conducted under high vacuum (< 1×10$^{-3}$ Pa) inside a cryostat (Janis VPF-800) with a coldfinger and a custom-made sample mount. As shown in Fig. 3(a), a test sample along with a heat flux sensor (FluxTeq, PHFS-01) of ±5% accuracy and a polyimide thin-film heater (OMEGA Engineering, KHLVA-101/10-P) was first attached to the 5-mm-thick acrylic carrier plate all in 1-inch-squared size. A thermistor (Mouser Electronics, SC30F103VN) with an accuracy of ±0.1°C was buried in the thermal paste between the sample and the heat flux sensor for measuring the sample temperature. After the acrylic carrier plate was pinned onto the brackets with about 2-mm spacing between the sample and the coldfinger, the cryostat was then brought down to high vacuum followed by liquid nitrogen (LN2) filling to cool down the coldfinger to 80 K, which mimics the cold space thermal environment. While deep space is colder at 3 K, radiative heat transfer from a body around 300 K to a heat sink at either 80 K or 3 K is almost the same with only 0.5% difference due to ~$T^4$ dependence.

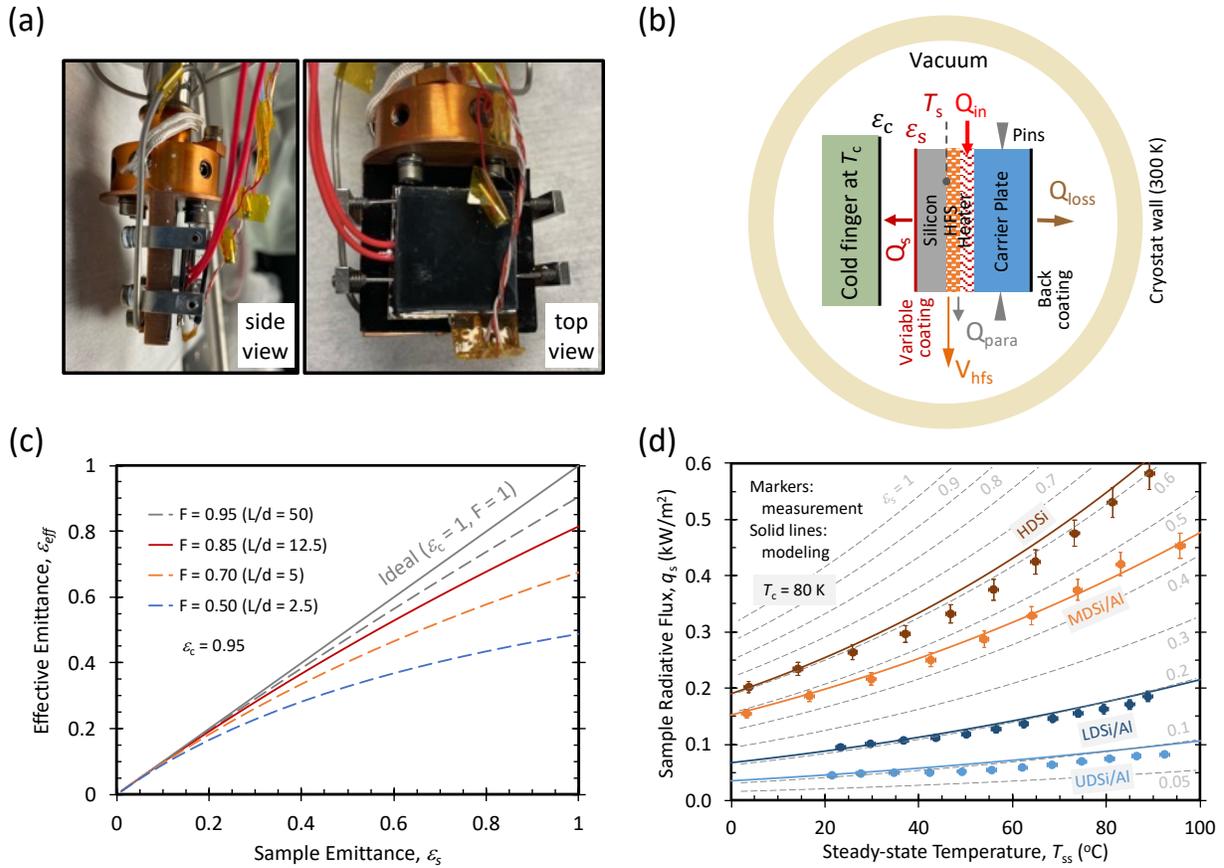

**Fig. 3.** (a) photos of sample mounting in the cryothermal setup; (b) heat transfer model for the cryothermal test; (c) calculated effective emittance ($\varepsilon_{eff}$) as a function of sample emittance ($\varepsilon_s$) for different view factor (*F*) values with the emittance ($\varepsilon_c$) of the cold finger covered by the black Actar taken as 0.95; and (d) validation on radiative heat flux at different steady-state temperatures from several reference samples with different static emittance values to the coldfinger at 80 K between cryothermal measurements and theoretical modeling.

During the cooling down stage, the sample temperature was maintained at 20°C by the heater with PID feedback control via a custom LabVIEW program, which also acquires the experimental data such as sample temperature, heater power, and heat flux voltage every one second with digital multimeters and programable power supply (Keithley 2000, 2100, 2200). At least 60 mins were waited after initial filling of LN2 for the system to stabilize thermally, and the LN2 was refilled full every two hours to ensure the stable coldfinger temperature. Once the test starts, the PID control was switched off and constant power input was used for the heater. For each given heater power, at least 20 mins were waited for the sample to reach steady state. The steady-state sample temperature was taken as the average of 300 data points over the last 5 mins with the standard deviation less than 1°C. To change the coldfinger temperature from 80 K to 25°C for simulating the internal radiative scenarios in space, a cryogenic temperature controller (Lakeshore 335) was used for the built-in heater on the coldfinger. After the tests with multiple heater power inputs were done, the sample heater was switched back to PID control for maintaining the sample at 20°C, which eventually boils off the remaining LN2 to bring the coldfinger back to room temperature for venting the cryostat and retrieving the sample.

Fig. 3(b) depicts the thermal model for the cryothermal test, where it involves multiple heat transfer interactions such as radiative heat transfer from the test sample to the coldfinger $Q_s$, heater power input $Q_{in} = VI$ with voltage and current from the power supply, parasitic heat transfer $Q_{para}$ due to conduction via pins and wires, as well as radiative heat loss $Q_{loss}$ from the backside to the cryostat wall. Note that, black Actar with near-unity emissivity of 0.951 was attached to the coldfinger for promoting heat absorption and to the backside of sample mount for ensuring steady state to be reached within 20 mins (see Fig. S2 for transient thermal modeling with different carrier plate materials and back coating materials). The energy on the sample mount is balanced at steady state as

$$Q_{in} = Q_s + Q_{para} + Q_{loss} \qquad (1)$$

Experimental radiative heat transfer from the sample to the coldfinger was found based on the measured heat flux sensor voltage $V_{hfs}$ and sample temperature $T_s$ as below:

$$Q_{s,exp} = \frac{V_{hfs}}{S(T_s)} - Q_{para}(T_s) \qquad (2)$$

Unknown heat flux sensor sensitivity $S(T_s)$ and parasitic heat load $Q_{para}(T_s)$ were first calibrated at a wide range of sample temperatures with black Actar and tungsten mirror by taking the sample radiative heat transfer from theory as

$$Q_{s,theo} = \varepsilon_{\text{eff}} \sigma A_c (T_s^4 - T_c^4) \qquad (3)$$

where $\varepsilon_{\text{eff}} = \left(\frac{1}{\varepsilon_s} + \frac{1}{F} + \frac{1}{\varepsilon_c} - 2\right)^{-1}$ is the effective emittance[34] depending on sample total hemispherical emittance $\varepsilon_s$, view factor $F$ between the sample and the coldfinger, and total emittance of the black-coated coldfinger $\varepsilon_c$= 0.951. Fig. 3(c) calculates the effective emittance $\varepsilon_{\text{eff}}$ as a function of sample emittance $\varepsilon_s$ with different view factor F values, which is determined based on the sample size-to-spacing ratio (*L/d*).[34] It can be seen that, the effective emittance could be much smaller than the sample emittance with non-unity view factors in particular for highly emissive samples. During the tests, the spacing between the sample and coldfinger was maintained at $d$ = 2±0.5 mm and the view factor was estimated to be $F$ = 0.85±0.03, at which the effective emittance of a black emitter decreases to 0.8.

Calibration tests with black Actar and tungsten mirror were carried out first with multiple heater power inputs to reach steady state temperatures between 0°C and 100°C (see Fig. S3 for the transient temperature profiles along with heater power and heat flux sensor voltage). Then the heat flux sensor sensitivity $S(T_s)$ was successfully fitted with a linear relation and the parasitic heat loss $Q_{para}(T_s)$ was also obtained as a function of sample temperature (see Fig. S4), both of which were used to find experimental sample radiative heat transfer $Q_{s,exp}$ based on the measured heat flux sensor voltage $V_{hfs}$ and steady-state temperature $T_{ss}$ for other static emitter samples and tunable VO2FP emitter.

To validate the cryothermal test with 80 K coldfinger as external radiative scenario in space, Fig. 3(d) presents the measured radiative heat flux from four static emitters of UDSi/Al, LDSi/Al, MDSi/Al and HDSi with different emittance values, which are in excellent agreement with the theoretical prediction. Note that ±5% uncertainty was considered for measured sample heat flux while ±1°C was given for measured steady-state temperatures. In particular, for the MDSi/Al sample with total emittance of 0.505, further heat transfer analysis in Fig. S5 reveals that, the sample radiative heat transfer $Q_s$ to the 80 K coldfinger is only ~35% of total heat input $Q_{in}$, whereas parasitic heat load $Q_{para}$ and backside loss $Q_{loss}$ respectively account for ~15% and ~50%. Note that the backside loss can be significantly reduced by using highly-reflective aluminum foil instead of black Actar, but this would lead to much longer time of about 5 times to reach steady state (see Fig. S2 for the transient thermal modeling).

Fig. 4(a) shows the variable radiative heat flux of the tunable VO2FP emitter at different steady-state temperatures to the coldfinger at 80 K for external radiative scenario in space observed from the cryothermal tests. The experimental data match well with theoretical modeling. Upon VO$_2$ phase transition, the radiative heat dissipation increases drastically from ~0.05 kW/m² at 45°C to ~0.45 kW/m² at 80°C. The radiative heat conductance, $G = \partial q_s / \partial T_{ss}$, is increased by nearly 6 times from 1.1 W/(m²·K) with insulating VO$_2$ below 45°C to 6.3 W/(m²·K) with metallic $VO_2$ above 80°C. During the phase transition, the radiative heat conductance is even higher with a value of 15.6 W/(m²·K) upon heating or 14.3 W/(m²·K) upon cooling.

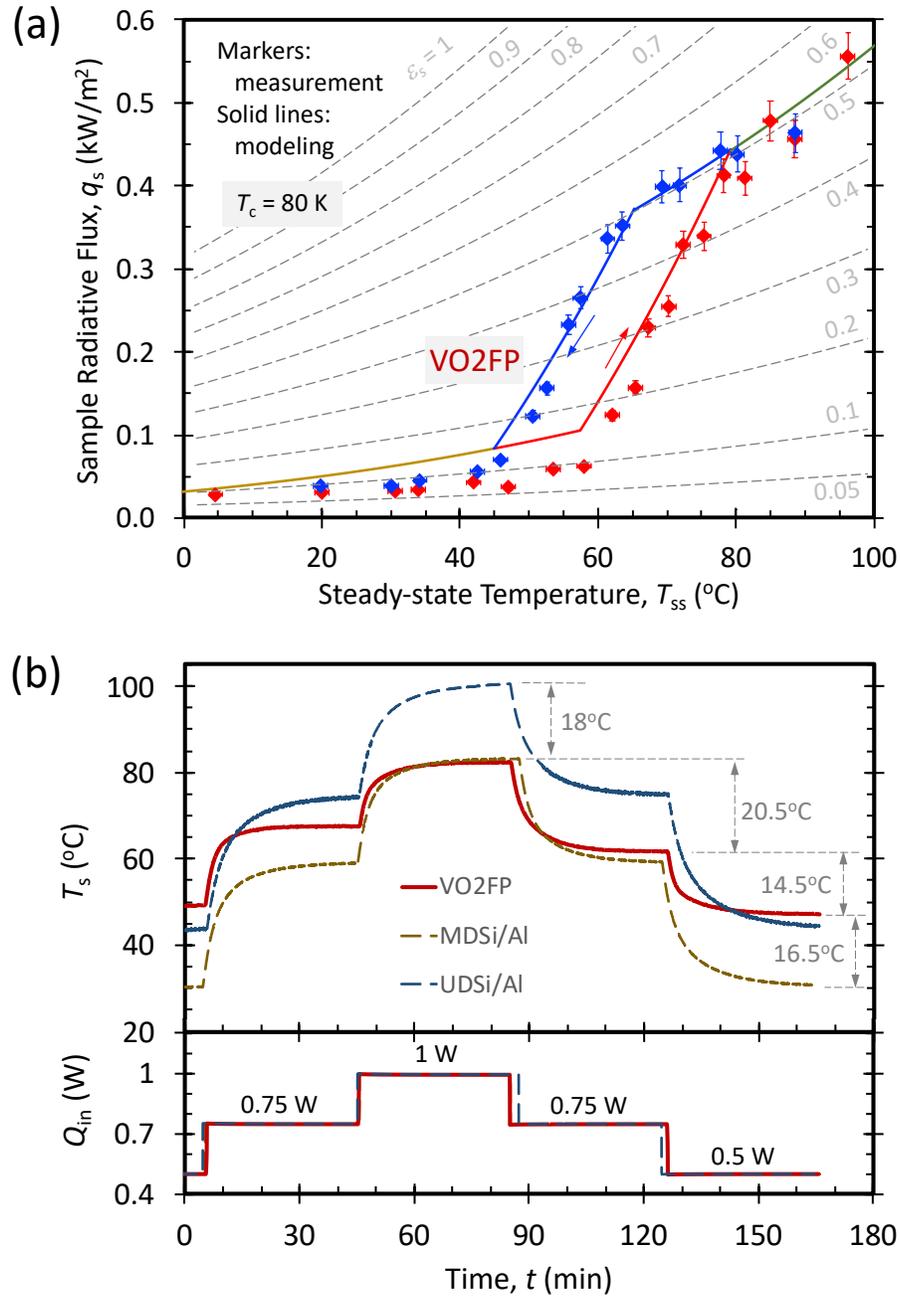

**Fig. 4.** Experimental demonstration of (a) variable radiative heat dissipation from the tunable VO2FP emitter at different steady-state temperatures in comparison with the theoretical modeling, and (b) thermal homeostasis with reduced temperature swing by tunable VO2FP emitter compared to static emitters (UDSi/Al and MDSi/Al) with step-wise heater power inputs. Note that the cold finger is maintained at 80 K for external radiative scenarios in space.

To experimentally demonstrate the thermal homeostasis effect, the tunable VO2FP emitter was heated in sequence with step-wise power inputs of 0.75 W, 1 W, 0.75 W and 0.5 W, and each power input was maintained for 40 mins. The static MDSi/Al ($\varepsilon_s$ = 0.505) and UDSi/Al ($\varepsilon_s$ = 0.105) samples, which have almost the same emittance of the tunable VO2FP emitter in its metallic and insulating phases, were also tested under the same heater power inputs for direct comparison. As shown in Fig. 4(b), when heated at 0.75 W, the VO2FP emitter reaches a steady temperature of 67.5°C, which is 8.5°C higher than MDSi/Al and 7°C lower than UDSi/Al. Note that at 67.5°C during heating, the VO2FP emitter just starts the phase transition, and its total hemispherical emittance is 0.315. Upon further heating at 1 W, the VO2FP emitter completes its phase transition into metallic phase with a total hemispherical emittance of 0.515, and it stabilizes at 82.3°C, which is almost the same as MDSi/Al but 18°C lower than the UDSi/Al.

With the heater power decreases to 0.75 W, the VO2FP emitter cools down nearly the same way as the MDSi/Al emitter to 61.8°C, where it has a total hemispherical emittance of 0.455. In the meantime, UDSi/Al is about 13°C hotter. Upon further cooling with 0.5 W heater power, the VO2FP emitter completes its transition into insulating phase with a much-decreased total hemispherical emissivity of 0.155. As a result, it has almost the same steady-state temperature with UDSi/Al instead around 47.3°C, which is 16.5°C higher than the MDSi/Al. From 1 W to 0.5 W heater power change, the vriable-emittance VO2FP coating experiences 35°C temperature swing, which is 21°C less than the static UDSi/Al sample and 18.5°C less than the MDSi/Al sample, demonstrating the thermal homeostasis effect. Note that about half of heater power is lost from the backside and ~15% is lost via pins and wires. With less heat loss such as using the high-reflective aluminum foil for the sample mount backing, the thermal homeostasis effect with smaller temperature swing is expected.

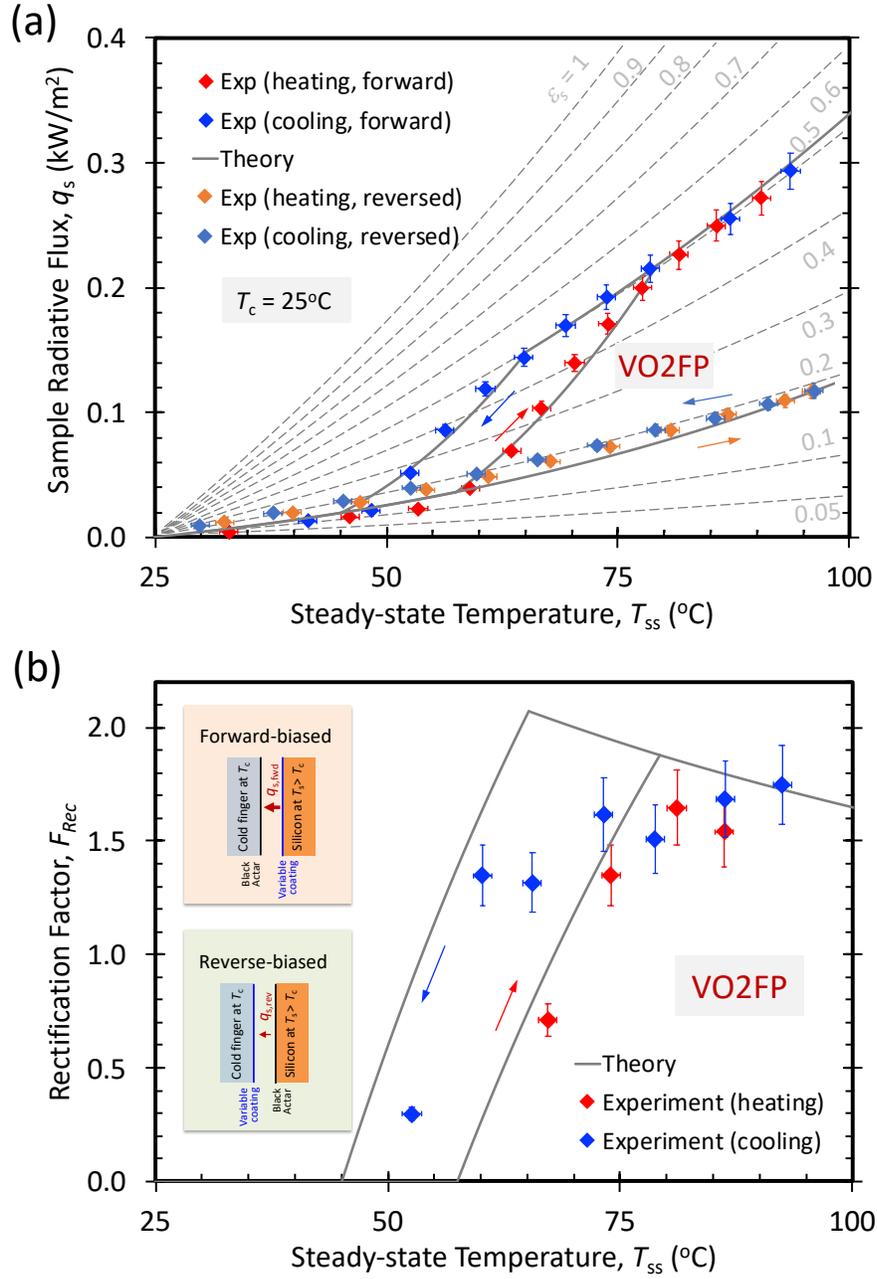

**Fig. 5.** Experimental demonstration of radiative thermal rectification with tunable VO2FP emitter: (a) measured radiative heat flux for the forward-biased and reversed-biased cases in comparison with theoretical modeling; (b) rectification factor ($F_{Rec}$) from both experiment and theoretical modeling. Inset illustrates the forward-biased case where the VO2FP emitter is heated and black Actar is placed on the cold finger maintained at 25°C and vice versa for the reversed-biased case as internal radiative scenarios in space.

To experimentally study the radiative heat transfer for the internal radiative scenarios in space, the cryostat coldfinger was heated to 25°C to mimic the internal thermal environment of the spacecraft. Fig. S6 shows the validation of cryothermal tests with multiple static emitters to the 25°C coldfinger, where excellent agreement is seen between the measurements and modeling. Fig. 5(a) presents the measured radiative heat flux when the tunable VO2FP emitter was heated from 25°C to 100°C dissipating heat radiatively to the black Actar attached on the 25°C coldfinger as the forward-biased scenario. Similar variable heat transfer with about 6-fold heat dissipation enhancement upon VO$_2$ phase transition is also observed, where the thermal conductance is improved from 0.9 W/(m$^2$·K) with insulating VO$_2$ below 45°C to 6.1 W/(m$^2$·K) with metallic VO$_2$ above 80°C, thanks to the variable emittance from the tunable VO2FP coating. During the phase transition, the thermal conductance reaches 8.4 W/(m$^2$·K) during heating and 6.4 W/(m$^2$·K) during cooling. However, when the VO2FP emitter is kept at 25°C on the coldfinger with the black Actar temperature varied, the radiative heat flux shows a nearly constant thermal conductance of 1.8 W/(m$^2$·K) for this reversed-biased scenario, clearly demonstrating much suppressed heat transfer at elevated temperatures with thermal rectification effect. The rectification factor, defined as $F_{Rec} = \frac{q_{s,fwd}}{q_{s,rev}} - 1$ where $q_{s,fwd}$ and $q_{s,rev}$ are respectively the radiative heat flux in forward- and reverse-biased scenarios,[7] could achieve up to 1.8±0.2 from the measurement with this tunable VO2FP emitter as shown in Fig. 5(b), which is in excellent agreement with theoretical modeling.

In summary, we have experimentally demonstrated dynamic heat control with thermal homeostasis and thermal recitation effects by the variable-emittance VO2FP coating in space-like environment mimicked inside a cryostat with a temperature-controlled coldfinger and a custom-made sample mount. Radiative heat dissipation was enhanced by 6 times in the thermal conductance upon VO$_2$ phase transition to both 80 K or 25°C coldfinger. For the external radiative scenario in space, reduced temperature swing by about 20°C was experimentally achieved with the tunable VO2FP coating compared to the static ones of the nearly same emittance. For the internal radiative scenario in space, suppressed radiative heat transfer was directly observed from the hot thermal background to the tunable coating maintained at 25°C with a thermal rectification factor of 1.8. Further with doping[35,36,37] to lower its phase transition temperatures, these tunable VO2FP coatings could play a crucial role in efficient passive thermal control of spacecraft in a given thermal environment for a particular space mission.

**Conflict of Interest**

The authors have no conflicts to disclose.

**Author Contributions**

L.W. conceived the idea, secured funds, supervised the project, characterized the samples, conducted the cryothermal tests, analyzed the data, prepared the Figures, and wrote the manuscript; N.B. set up the cryothermal apparatus and performed preliminary test; S.T. fabricated the samples; C.S. repeated the cryothermal tests for validation; all authors reviewed the final manuscript.

**Acknowledgements**

This work was supported by National Science Foundation (CBET-2212342). S.T. would thank the PhD Fellowship support from National Aeronautics and Space Administration (NNX16AM63H). N.B and C.S are grateful to the supports from the Master's Opportunity for Research in Engineering (MORE) program, the Fulton Undergraduate Research Initiative (FURI) program, and Barrett, The Honors College at Arizona State University.

**Data Availability**

The data that support the findings of this study are available from the corresponding author upon reasonable request.

**References**

1. V. L. Pisacane, "Spacecraft Systems Design and Engineering," Encyclopedia of Physical Science and Technology, 3rd Ed., 464-483 (Elsevier, 2003).
2. R. D. Karam, "Satellite Thermal Control for Systems Engineers," Progress in Astronautics and Aeronautics, 181 (1998).
3. D. W. Hengeveld, M. M. Mathison, J. E. Braun, E. A. Groll, and A. D. Williams, *HVAC&R Research*, 16, 189–220 (2010)

4. T. Swanson, B. Motil, F. Chandler, W. Bruce, C. Dinsmore, C. Kostyk, M. Lysek, S. Rickman, and R. Stephan, "NASA Technology Roadmaps TA 14: Thermal Management Systems," National Aeronautics and Space Administration, Washington, DC, 21 (2015).
5. F. Lang, H. Wang, S. Zhang, J. Liu, and H. Yan, *Int. J. Thermophys.* 39, 6 (2016).
6. Wu S.-H., Chen M., Barako M. T., Jankovic V., Hon P. W. C., Sweatlock L. A. and Povinelli M. L., *Optica* 4, 1390–1396 (2017).
7. Wang, L.P. and Zhang, Z.M., *Nanosc. Microsc. Thermophys. Eng. 17*, 337-348 (2013).
8. Shimakawa Y., Yoshitake T., Kubo Y., Machida T., Shinagawa K., Okamoto A., Nakamura Y., Ochi A., Tachikawa S. and Ohnishi A., *Appl. Phys. Lett.* 80, 4864–4866 (2002).
9. Fan D., Li Q. and Dai P., *Acta Astronaut.* 121, 144–152 (2016).
10. Benkahoul M., Chaker M., Margot J., Haddad E., Kruzelecky R., Wong B., Jamroz W. and Poinas P., *Sol. Energy Mater. Sol. Cells* 95, 3504–3508 (2011).
11. Hendaoui A., Émond N., Dorval S., Chaker M. and Haddad E., *Sol. Energy Mater. Sol. Cells* 117, 494–498 (2013).
12. Fan D., Li Q., Xuan Y. and Dai P., *Thin Solid Films* 570, 123–128 (2014).
13. Taylor S., Yang Y. and Wang L., *J. Quant. Spectrosc. Radiat. Transf.* 197, 76–83 (2017).
14. Shrewsbury B. K., Morsy A. M. and Povinelli M. L., *Opt. Mater. Express* 12, 1442-1449 (2022).
15. Shrewsbury, B.K., Yu, R., Barako, M.T., Lien, M.R., Rosenzweig, R., Howes, A. and Povinelli, M.L., *Opt. Express 32*, 43430-43444 (2024).
16. Hendaoui A., Émond N., Chaker M. and Haddad É., *Appl. Phys. Lett.* 102, 061107 (2013).
17. Wang X., Cao Y., Zhang Y., Yan L. and Li Y., *Appl. Surf. Sci.* 344, 230–235 (2015).
18. Cesarini G., Leahu G., Li Voti R. and Sibilia C., *Infrared Phys. Tech.* 93, 112–115 (2018).
19. Kim H., Cheung K., Auyeung R. C. Y., Wilson D. E., Charipar K. M., Piqué A. and Charipar N. A., *Sci. Rep.* 9, 11329 (2019).
20. Taylor S., Long L., Mcburney R., Sabbaghi P., Chao J. and Wang L., *Sol. Energy Mater. Sol. Cells* 217, 110739 (2020).
21. Yu, R., Shrewsbury, B.K., Wu, C., Kumarasubramanian, H., Surendran, M., Ravichandran, J. and Povinelli, M.L., *Appl. Phys. Lett.* 125, 121117 (2024).

22. Kort-Kamp W. J. M., Kramadhati S., Azad A. K., Reiten M. T. and Dalvit D. A. R., *ACS Photon.* 5, 4554–4560 (2018).

23. Sun K., Riedel C. A., Urbani A., Simeoni M., Mengali S., Zalkovskij M., Bilenberg B., De Groot C. H. and Muskens O. L., *ACS Photon.* 5, 2280–2286 (2018).

24. Chen M., Morsy A. M. and Povinelli M. L., *Opt. Express* 27, 21787–21793 (2019).

25. Numan N., Mabakachaba B., Simo A., Nuru Z. and Maaza M., *JOSA A* 37, C45–C49 (2020).

26. Long L., Taylor S. and Wang L., *ACS Photon.* 7, 2219–2227 (2020).

27. Araki K. and Zhang R. Z., *AIP Adv.* 12, 055205 (2022).

28. Barako M. T., Howes A., Sweatlock L. A., Jankovic V., Hon P. W. C., Tice J., Povinelli M. and Knight M. W., *J. Thermophysics Heat Trans.* 36, 1003-1014 (2022).

29. Dong K., Tseng D., Li J., Warkander S., Yao J. and Wu J., *Cell Rep. Phys. Sci.* 3, 101066 (2022).

30. Chen, Y., Zhao, T., Geng, C., Chang, Y., Lu, J., Zhao, Q., Chen, Y., Ouyang, H., Dou, S., Gu, J. and Li, Y., *Results Eng.* 27, 106262 (2025).

31. Morsy A. M., Barako M. T., Jankovic V., Wheeler V. D., Knight M. W., Papadakis G. T., Sweatlock L. A., Hon P. W. C. and Povinelli M. L, *Sci. Rep.* 10, 13964 (2020).

32. Taylor, S., Boman, N., Chao, J. and Wang, L., *Appl. Therm. Eng.* 199, 117561 (2021).

33. Handbook of Optical Constants of Solids, edited by E. D. Palik (Academic, San Diego, 1998).

34. Fundamentals of Heat and Mass Transfer, F. P. Incropera, D. P. DeWitt, T. L. Bergman, and A. S. Lavine (Wiley, New York, 1996).

35. T. D. Manning, I. P. Parkin, M. E. Pemble, D. Sheel and D. Vernardou, *Chem. Mater.* 16, 744–749 (2004).

36. J. Rensberg, S. Zhang, Y. Zhou, A. S. McLeod, C. Schwarz, M. Goldflam, M. Liu, J. Kerbusch, R. Nawrodt, S. Ramanathan, D. N. Basov, F. Capasso, C. Ronning, and M. A. Kats, *Nano Lett.* 16, 1050–1055 (2016).

37. V. K. Rajan, J. Chao, S. Taylor, and L. Wang, *J. Appl. Phys.* 137, 195302 (2025).

# Supplementary Material

**Reducing Temperature Swing and Rectifying Radiative Heat Transfer for Passive Dynamic Space Thermal Control with Variable-Emittance Coatings**

Liping Wang,* Neal Boman, Sydney Taylor, and Chloe Stoops

School for Engineering of Matter, Energy and Transport, *Arizona State University, Tempe, Arizona 85287*

* Corresponding author: liping.wang@asu.edu

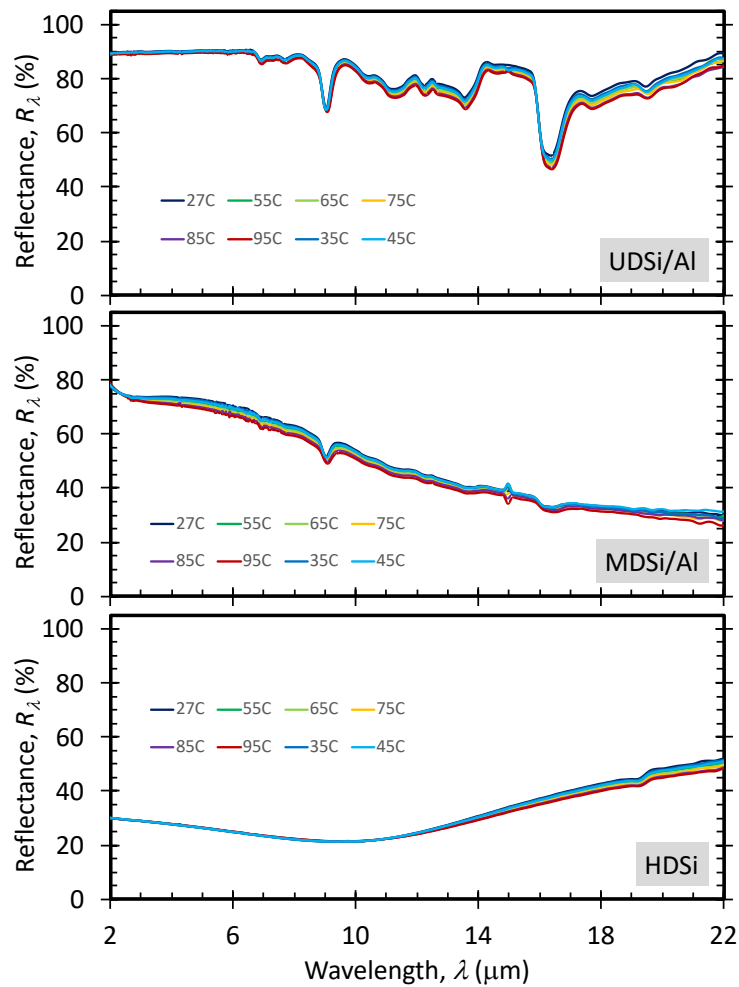

**Fig. S1.** Measured temperature-dependent infrared spectral reflectance of lightly-doped, medium-doped and heavily-doped silicon wafers of 280-μm thick with 200-nm Al backside coating from 27°C to 95°C, where negligible temperature effect is observed.

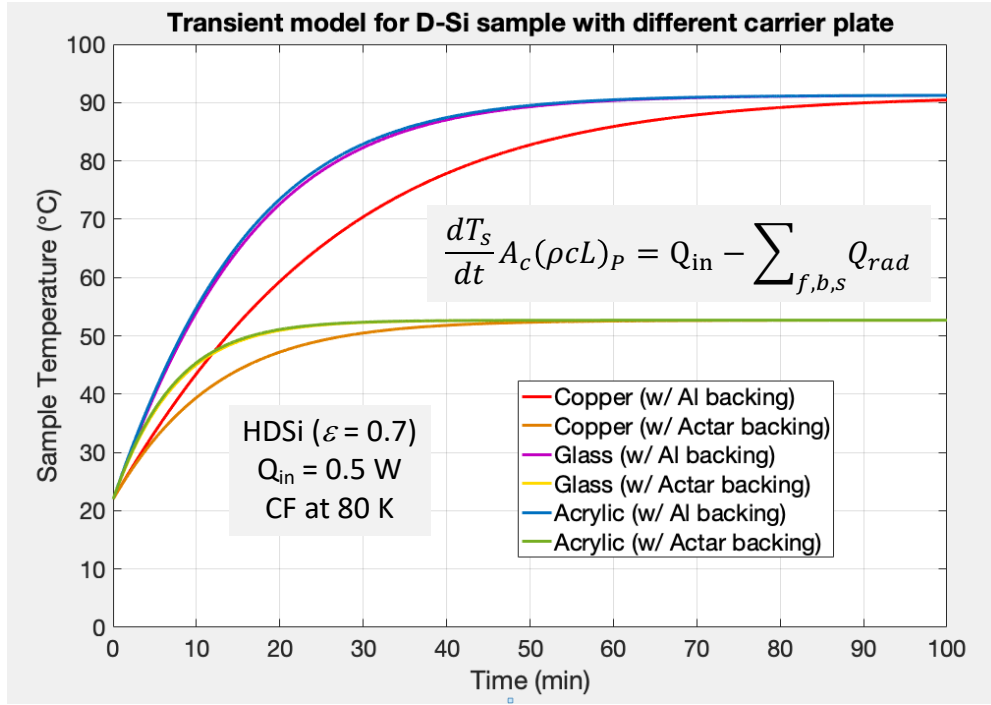

**Fig. S2.** Analytical modeling of transient temperature change of heavily doped silicon (emittance = 0.7) in radiative heat exchange to the black cold finger at 80 K under 0.5 W heater power input when mounted on the 5-mm-thick sample carrier plate made of copper, glass or acrylic with either black Actar or reflective Al backing. Steady state is reached quickest by using the acrylic carrier plate with black Actar backing (~20 mins) used in this work, which is about 5 times faster than copper with Al backing (~100 mins) used in our previous work.

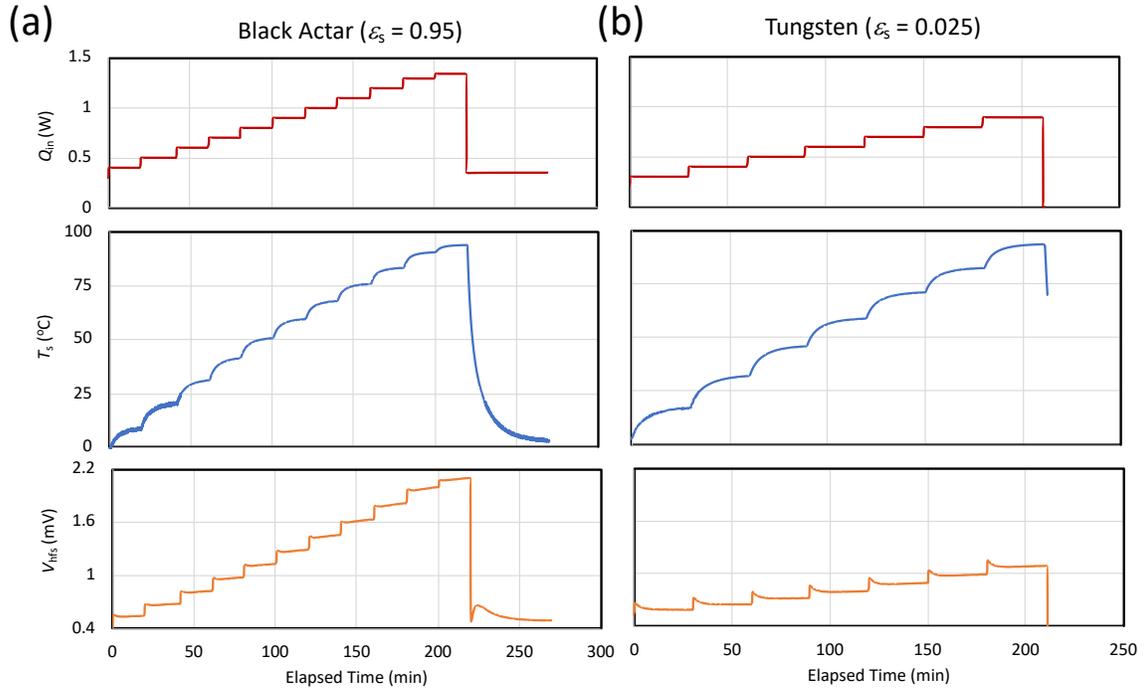

**Fig. S3.** Heater power input, sample temperature and voltage reading from heat flux sensor as a function of time from the cryothermal tests for (a) the black Actar ($\varepsilon_s$=0.95) and (b) the tungsten mirror ($\varepsilon_s$=0.025), whose steady-state data are used for calibrating the temperature-dependent sensitivity $S(T)$ and parasitic heat loss $Q_{para}(T)$ for the heat flux sensor.

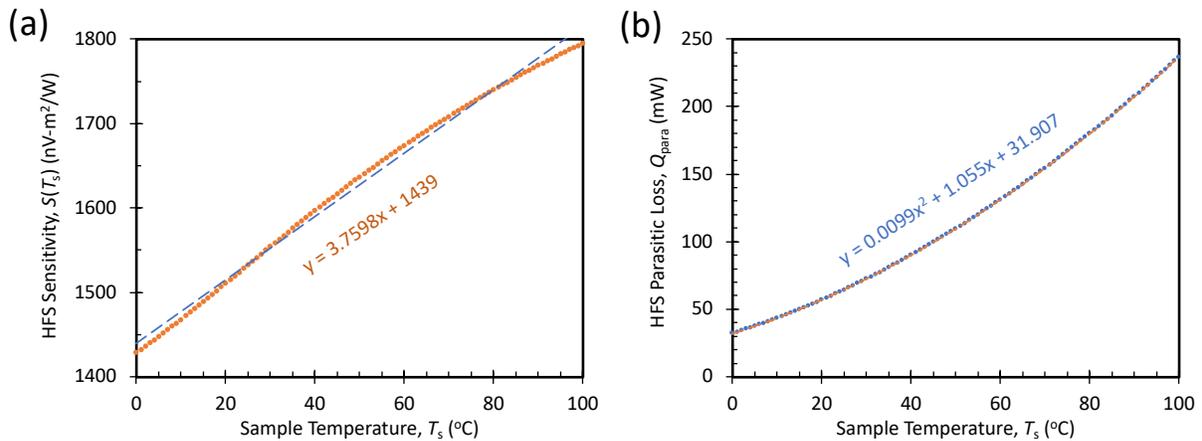

**Fig. S4.** Calibrated (a) sensitivity $S(T)$ and (b) parasitic heat loss $Q_{para}(T)$ for the heat flux sensor as a function of sample temperature based on the cryothermal tests for the black Actar and tungsten mirror samples. Polynomial fitting is used for both to process the HFS voltage and temperature readings from the cryothermal tests to obtain the experimental radiative heat flux for other samples including the tunable VO2FP emitter and other static emitters.

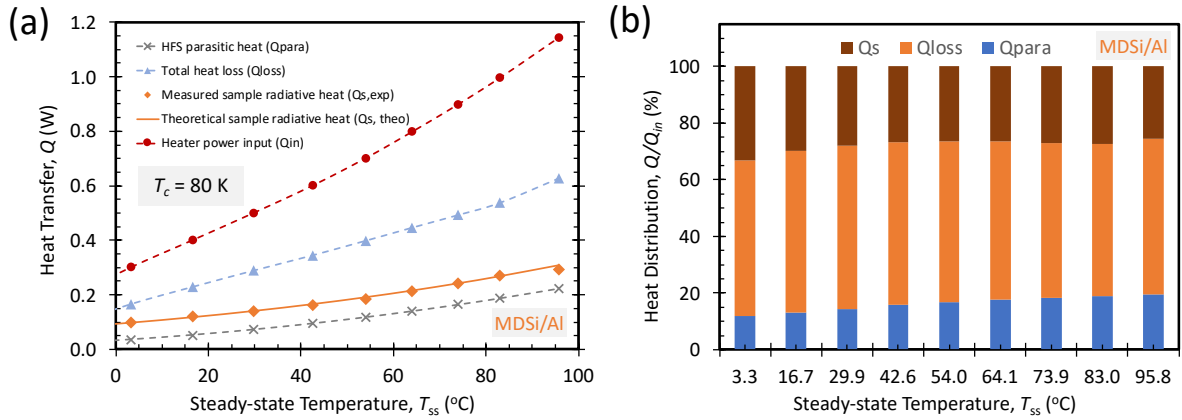

**Fig. S5.** Heat transfer analysis of the cryothermal test for the medium doped silicon with Al backing (MDSi/Al) sample ($\varepsilon_s$=0.51): (a) absolute value of each heat transfer mode including heater power ($Q_{in}$), parasitic heat loss from the heat flux sensor ($Q_{para}$), total heat loss ($Q_{loss}$), measured sample radiative heat flux ($Q_{s,exp}$), theoretical sample radiative heat flux ($Q_{s,theo}$); (b) heat distribution of each heat transfer normalized to the heater power input.

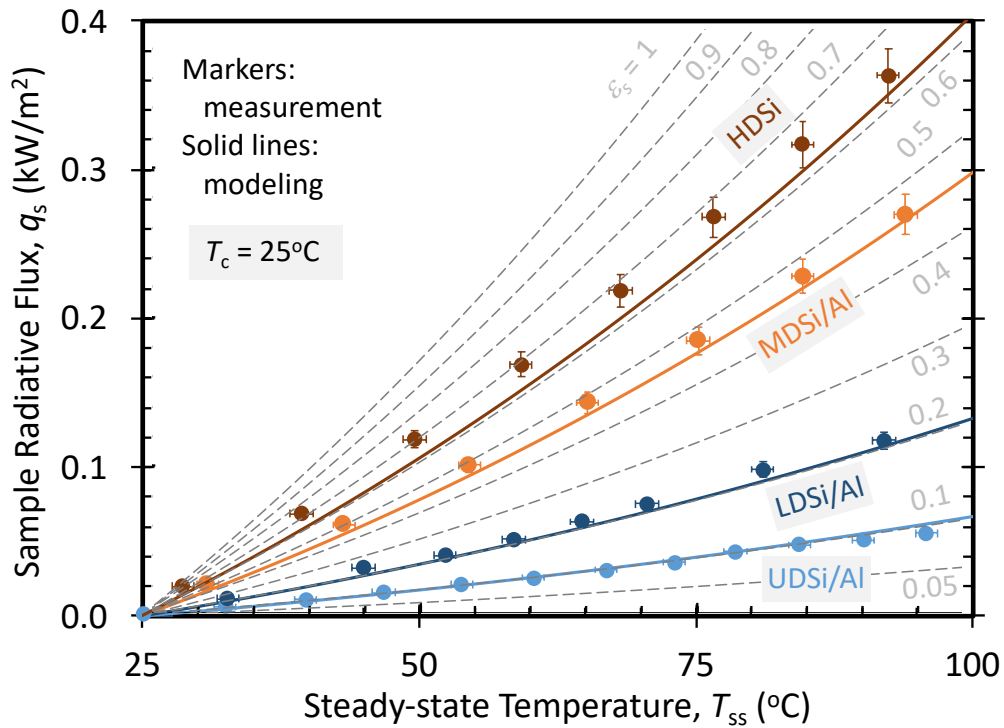

**Fig. S6.** Validation on radiative heat flux at different steady-state temperatures from several reference samples with different static emittance values to the cold finger at 25°C between cryothermal measurements and theoretical modeling.